\documentclass{physeauth}
\usepackage{graphicx}
\usepackage{amsmath}
\usepackage{amssymb}

\begin{document}

\begin{frontmatter}

\title{Coexistence of full which-path information and interference in Wheeler's delayed choice experiment with photons}

\author[embd]{K. Michielsen},
\author[rug]{S. Yuan},
\author[rug]{S. Zhao},
\author[rug]{F. Jin},
\author[rug]{H. De Raedt\thanksref{thank1}}

\address[embd]{EMBD, Vlasakker 21, 2160 Wommelgem, Belgium}
\address[rug]{Department of Applied Physics, Zernike Institute for Advanced Materials,
University of Groningen, Nijenborgh 4, 9747 AG Groningen, The Netherlands}

\thanks[thank1]{
Corresponding author.
E-mail: h.a.de.raedt@rug.nl}

\begin{abstract}
We present a computer simulation model
that is a one-to-one copy of an experimental realization of Wheeler's delayed choice experiment
that employs a single photon source and a Mach-Zehnder interferometer composed of a 50/50 input beam splitter and
a variable output beam splitter with adjustable reflection coefficient $R$ (V. Jacques {\sl et al.}, Phys. Rev. Lett. 100, 220402 (2008)).
For $0\le R\le 0.5$, experimentally measured values of the interference visibility $V$ and the path
distinguishability $D$, a parameter quantifying the which-path information WPI, are found to fulfill the complementary relation
$V^2+D^2\le 1$, thereby allowing to obtain partial WPI while keeping interference with limited visibility.
The simulation model that is solely based on experimental facts,
that satisfies Einstein's criterion of local causality and that does not rely on any concept of quantum theory or of probability theory,
reproduces quantitatively the averages calculated from quantum theory.
Our results prove that it is possible to give a particle-only description of the experiment, that one can have full WPI
even if $D=0$, $V=1$ and therefore that the relation $V^2+D^2\le 1$ cannot be regarded as quantifying the notion of complementarity.
\end{abstract}


\end{frontmatter}

\section{Introduction}

Particle-wave duality, a concept of quantum theory, attributes to photons the
properties of both wave and particle behavior depending upon the circumstances of
the experiment~\cite{HOME97}.
The particle behavior of photons has been shown in an experiment composed of a
single beam splitter (BS) and a source emitting single photons and pairs of
photons~\cite{GRAN86}. The wave character has been demonstrated in a single-photon Mach-Zehnder
interferometer (MZI) experiment~\cite{GRAN86}.
In 1978, Wheeler proposed a gedanken
experiment~\cite{WHEE83}, a variation on Young's double slit experiment, in which the
decision to observe wave or particle behavior is made after the photon has
passed the slits. The pictorial description of this experiment defies common sense:
The behavior of the photon in the past is said to be changing from a particle to a wave or vice versa.

\begin{figure*}[t]
\begin{center}
\includegraphics[width=15cm]{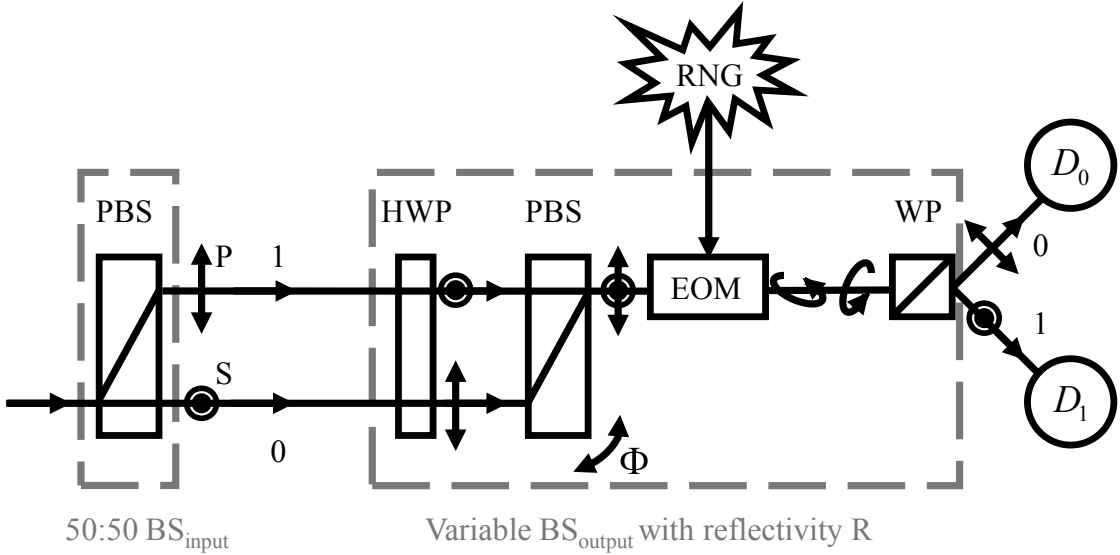}
\caption{Schematic diagram of the experimental setup for Wheeler's delayed-choice gedanken
experiment~\cite{JACQ08}. PBS: Polarizing beam splitter; HWP:
Half-wave plate; EOM: electro-optic modulator; RNG: Random number generator;
WP: Wollaston prism; P,S: Polarization state of the photon; $\Phi$: Phase shift between paths 0 and 1; $D_{0}$, $D_{1}$: Detectors.}
\label{wheeler}
\end{center}
\end{figure*}

Recently, Jacques {\sl et al.}
reported on an experimental realization of Wheeler's delayed choice experiment using
a single photon source and a MZI composed of a 50/50 input BS and a variable output
BS with adjustable reflection coefficient $R$~\cite{JACQ08}, a modification of the experiment presented in \cite{JACQ07} in
which two 50/50 BSs were used. A schematic picture of
the experimental set-up is shown in Fig.~\ref{wheeler}. The reflection coefficient $R$
of the variable beam splitter (BS$_{output}$) can be controlled by a
voltage applied to an electro-optic modulator (EOM), making it act as a variable wave plate.
This can be done after each photon has entered the MZI.
The phase-shift $\Phi$ between the two arms of the MZI is varied by tilting the polarizing beam splitter (PBS)
of the variable output BS.
For $0\leq R\leq 0.5$ measured values of the interference
visibility $V$~\cite{BORN64} and the path distinguishability $D$~\cite{JACQ08}, a parameter that quantifies the which-path information (WPI),
were found to fulfill the complementary relation $V^2+D^2\leq 1$.
The extreme situations ($V=0$, $D=1$) and ($V=1$, $D=0$), obtained for $R=0$ and $R=0.5$, give full and no WPI, associated
with particlelike and wavelike behavior, respectively.
By choosing $0<R<0.5$ Jacques {\sl et al.} claim to have obtained partial WPI while keeping interference with limited visibility~\cite{JACQ08}, thereby
having accomplished an affirmative delayed choice test of complementarity or wave-particle duality as it is often phrased.

Although the detection events (detector ``clicks'') are the only experimental facts,
the pictorial description of Jacques {\sl et al.}~\cite{JACQ08,JACQ07} is as follows:
Linearly polarized single photons are sent through a 50/50 PBS (BS$_{input}$),
spatially separating photons with S polarization (path 0) and P polarization (path 1) with equal frequencies.
After the photon has passed BS$_{input}$, but before the photon
enters the variable BS$_{output}$ the decision to apply a voltage to the EOM is made.
The PBS of BS$_{output}$ merges the paths of the orthogonally polarized photons travelling paths 0 and 1 of the MZI, but
afterwards the photons can still be unambiguously identified by their polarizations.
If no voltage is applied to the EOM then $R=0$ and the EOM can be regarded as doing nothing to the photons.
Because the polarization eigenstates of the Wollaston prism correspond to the P and S polarization
of the photons travelling path 0 an 1 of the MZI, each detection event
registered by one of the two detectors $D_0$ or $D_1$
is associated with a specific path (path 0 or 1, respectively).
Both detectors register an equal amount of detection events, independent of the phase shift $\Phi$ in the MZI.
This experimental setting, corresponding to the open configuration of the MZI, clearly gives full WPI about the photon within
the interferometer (particle behavior), characterized by $D=1$.
In this case no interference effects are observed, corresponding with a zero interference visibility ($V=0$).
When a voltage is applied to the EOM, then $R\ne 0$ (see Eq.~(2) in ~\cite{JACQ08}) and
the EOM acts as a wave plate rotating the polarization of the incoming photon by an angle depending on $R$.
The Wollaston prism partially recombines the polarization of the photons that have travelled along
different optical paths with phase difference $\Phi$ (closed configuration), and interference appears ($V\ne0$), a result
expected for a wave. The WPI is partially washed out, up to be totally erased when $R=0.5$ ($D=0$, $V=1$).

The outcome of delayed-choice experiments~\cite{JACQ08,JACQ07,HELL87,BALD89,LAWS96,KAWA98,KIM00}, that is the average results of many detection
events, is in agreement with wave theory (Maxwell or quantum theory).
However, the pictorial description explaining the experimental facts~\cite{JACQ07} defies common sense: The decision to apply a voltage
to the EOM after the photon left BS$_{input}$ but before it passes BS$_{output}$,
influences the behavior of the photon in the past and changes the
representation of the photon from a particle to a wave~\cite{JACQ07}.
Although on one hand quantum theory can be used to describe the final outcome of this type of experiments
(the average results of many detection events),
on the other hand it does not describe single events~\cite{HOME97}.
Therefore, it should not be a surprise that the application of concepts
of quantum theory to the domain of individual events may lead to conclusions
that are at odds with common sense.
Although not applying this reasoning to describe this type of experiments could prevent us from making nonsensical
conclusions, this unfortunately would not give us a single clue as how to explain the fact that individual events are observed
experimentally and, when collected over a sufficiently long time, yield averages that agree with quantum theory.

Since no theory seems to exist that can give a sensical description of the ``whole'' experiment,
we adopted the idea to search for algorithms that could mimic (simulate) the detection events
and experimental processes, including for example the random switching of the EOM for each photon sent into the interferometer.
We moreover require that the algorithms used to simulate the action of an optical element, such as a BS or wave plate, on a photon should
be independent of the experimental setup. In other words, the algorithms to simulate an optical component should be the same for all
identical optical components within the same experiment but also within a different experiment.
Hence, first solving the Schr\"odinger equation for a given experimental configuration and then simply generating events according
to the resulting probability distribution is not what we have in mind when we perform an event-by-event simulation of the experiment.
Similarly, first calculating the quantum potential (which requires the solution of the Schr\"odinger equation)
and then solving for the Bohm trajectories
is a different kind of event-by-event simulation than the one we describe in this paper.
In this paper, the event-by-event simulation algorithm reproduces the results of quantum theory,
without first solving a wave equation.

In this paper, we describe a model that, when implemented
as a computer program, performs an event-by-event simulation
of Wheeler's delayed-choice experiment.
Every essential component of the laboratory experiment
(PBS, EOM, HWP, Wollaston prism, detector) has a counterpart in the
algorithm.
The data is analyzed by counting detection events, just like in the experiment~\cite{JACQ08,JACQ07}.
The simulation model is solely based on experimental facts, satisfies Einstein's criterion
of local causality and does not rely on any concept of quantum theory or of probability theory.
Nevertheless, our simulation model reproduces the averages obtained from the quantum theoretical
description of Wheeler's delayed choice experiment but as our approach does
not rely on concepts of quantum theory and gives a description on the
level of individual events, it provides a description of the experimental facts that
does not defy common sense.
In a pictorial description of our simulation model, we may speak about ``photons'' generating
the detection events. However, these so-called photons, as we will call them in the sequel,
are  elements of a model or theory for the real laboratory experiment only.
The experimental facts are the settings of the various apparatuses and
the detection events. What happens in between activating the source and the registration of the detection
events is not measured and is therefore not known.
Although we always have full WPI of the photons
in the closed configuration of the interferometer (we can always track the photons during the simulation),
the photons build up an interference
pattern at the detector. The appearance of an interference pattern is
commonly considered to be characteristic for a wave. In this paper, we demonstrate
that, as in experiment, it can also be build up by many photons.
These photons have full WPI, never directly communicate with each other
and arrive one by one at a detector.

The work described in this paper elaborates on the work described in~\cite{ZHAO08a} to simulate the experiment
reported in~\cite{JACQ07}.
The simulation model is built on earlier work~\cite{RAED05d,RAED05b,RAED05c,MICH05,RAED06c,RAED07a,RAED07b,RAED07c,ZHAO08b,ZHAO08c} that
demonstrated that it may be possible to simulate quantum phenomena on the level of individual events
without invoking a single concept of quantum theory.
Specifically, we have demonstrated that locally-connected networks of processing units
with a primitive learning capability can simulate event-by-event,
the single-photon beam splitter and MZI experiments of Grangier {\sl et al.}~\cite{GRAN86}
and Einstein-Podolsky-Rosen experiments with photons~\cite{ASPE82a,ASPE82b,WEIH98}.
Furthermore, we have shown that this approach can be generalized
to simulate universal quantum computation by an event-by-event process~\cite{RAED05c,MICH05}.
Our event-by-event simulation approach rigorously satisfies
Einstein's criterion of local causality and builds up the final outcome that agrees
with quantum theory event-by-event, as observed in real experiments.

\section{Simulation model}

The simulation algorithm can be viewed as a message-processing and message-passing
process: It routes messengers one by one through a network of units that process messages.
The messengers may be regarded as ``particles''.
These messengers carry a message which contains information about for example the relative time the particle traveled,
its polarization, its color, its velocity, and so on.
In other words, the message represents a so-called variable property of the particle that can
be manipulated and measured given particular experimental setttings.
The components of the experimental setup
such as the BS, the wave plates, the Wollaston prism and so on are so-called processing units
that interpret and manipulate the messages carried by the particles. These processing
units are put in a network that represents the complete experimental setup.
Since at any given time there is only one messenger being routed through the whole network,
there is no direct communication between the messengers. The only form of communication
is through the processing units when the messengers are routed through the network.
The model satisfies the intuitive notion of local causality.

In general, processing units consist of an input stage,
a transformation stage, an output stage and have an internal vector representing their internal state.
The input (output) stage may have several channels
at (through) which messengers arrive (leave).
Some processing units are simpler in the sense that the input stage
is not necessary for the proper functioning of the device.
As a messenger arrives at an input channel of a processing unit,
the input stage updates its internal vector, and
sends the message, represented by a vector, together with its internal vector
to the transformation stage that implements the operation of the particular device.
Then, a new message is sent to the output stage,
using a pseudo-random number to select the output channel
through which the messenger will leave the unit.
We use pseudo-random numbers to mimic the apparent unpredictability of the
experimental data only.  The use of pseudo-random numbers is merely convenient, not essential.

In the experimental realization of Wheeler's delayed choice experiment by
Jacques {\sl et al.}~\cite{JACQ08} linearly polarized single photons
are sent through a PBS that
together with a second, movable, variable output PBS with adjustable reflectivity $R$ forms an interferometer (see Fig.~\ref{wheeler}).
The basic idea now is that we have to construct a model for the messengers representing the photons and
for the processing units representing the optical components in the experimental setup.
We require that the processing units for identical optical components should be reusable within the same and whitin different experiments.
The network of processing units is a one-to-one
image of the experimental setup~\cite{JACQ08} and is shown in Fig.~\ref{wheeler}.
In what follows we describe some elements of our model in more detail.
Additional information can be found in~\cite{ZHAO08a}.

\subsection{Messengers}

In a pictorial description of the experiment the photons can be
regarded as particles playing the role of messengers.
Each messenger carries a (variable) message which contains information about its
phase and polarization.
The phase combines information about the frequency of the light source and
the time that particles need to travel a given path.
However, no explicit information about distances and frequencies is required
since we can always work with relative phases.

\begin{figure*}[t]
\begin{center}
\includegraphics[width=10cm]{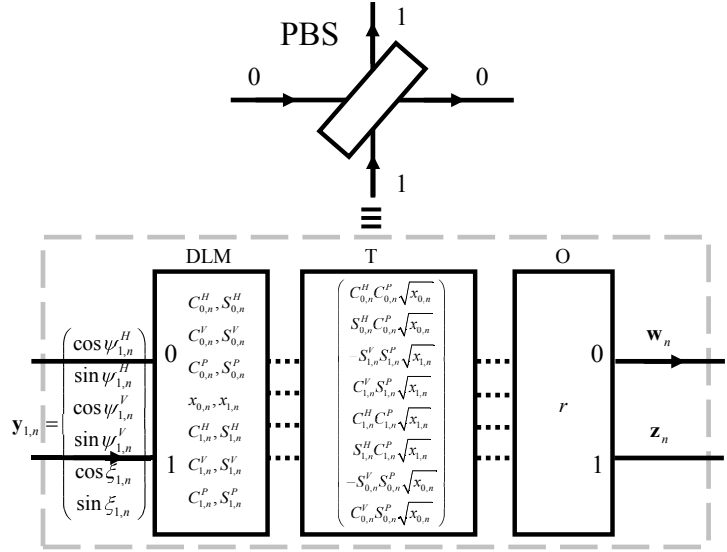}
\caption{Diagram of a DLM-based processing unit that performs an event-based simulation of a
polarizing beam splitter (PBS). The solid lines represent the
input and output channels of the PBS. The presence of a message is indicated
by an arrow on the corresponding channel line. The dashed lines indicate the
data flow within the PBS.}
\label{PBS}
\end{center}
\end{figure*}

The information carried by the messenger can be represented by a
six-dimensional unit vector ${\bf y}_{k,n}=( \cos \psi
_{k,n}^{H},\sin \psi _{k,n}^{H},\cos \psi _{k,n}^{V},\sin \psi
_{k,n}^{V},\cos \xi _{k,n},\sin \xi _{k,n}) $.
The superscript H (V) refers to the horizontal (vertical)
component of the polarization
and $\psi _{k,n}^{H}$, $\psi _{k,n}^{V}$, and $\xi _{k,n}$
represent the phases and polarization of the photon, respectively.
It is evident that the representation used here maps one-to-one
to the plane-wave description of a classical electromagnetic field~\cite{BORN64},
except that we assign these properties to each individual photon, not to a wave.
The subscript $n\geq 0$ numbers
the consecutive messages and $k=0,1$ labels the channel of the PBS
at which the message arrives (see below).

Since in this paper we will demonstrate explicitly that in our model photons always have full WPI even if
interference is observed, we give the messengers one extra label, the path label having the value 0 or 1.
The information contained in this label is not accessible in the experiment~\cite{JACQ08}.
We only use it to track the photons in the network. The path label is set in the input BS and remains
unchanged until detection. Therefore we do not consider this label in the description of the processing units
but take it into account when we detect the photons.

\subsection{Polarizing beam splitter}
From classical electrodynamics we know that if an electric field is applied to a dielectric material
the material becomes polarized~\cite{BORN64}. The polarization ${\bf P({\bf k},t)}$ is given by
\begin{equation}
{\bf P}({\bf k},t)=\int_0^t\chi({\bf k},u){\bf E}({\bf k},t-u) du,
\label{polarization}
\end{equation}
where ${\bf E}({\bf k},t)$ denotes the electric field vector,
${\bf k}$ is the wave vector, and $\chi$ is the linear response function~\cite{BORN64}.
From Eq.~(\ref{polarization}) it is evident
that the dielectric material shows some kind of memory effect because the response (the polarization)
of the material to the applied electric field is a function of both present and past values of the electric field.
We use this kind of memory effect in our algorithm to model the PBS.

The processor that performs the event-by-event simulation of a PBS is depicted in Fig.~\ref{PBS}.
It consists of an input stage, a simple deterministic learning machine (DLM)~\cite{RAED05d,RAED05b,RAED05c,MICH05},
a transformation stage (T), an output stage (O) and
has two input and two output channels labeled with $k=0,1$.
We now define the operation of each stage explicitly.

\begin{itemize}
\item{Input stage: The DLM receives a message on either input channel 0 or 1,
never on both channels simultaneously. The arrival of a message on channel 0
(1) is named a 0 (1) event. The input events are represented by the vectors
${\bf v}_{n}=(1,0)$ or ${\bf v}_{n}=(0,1)$ if the $n$th
event occurred on channel 0 or 1, respectively.
The DLM has six internal registers ${\bf Y}_{k,n}^{H}=(
C_{k,n}^{H},S_{k,n}^{H}) ,$ ${\bf Y}_{k,n}^{V}=(
C_{k,n}^{V},S_{k,n}^{V}) ,$ ${\bf Y}_{k,n}^{P}=(
C_{k,n}^{P},S_{k,n}^{P}) $ and one internal vector ${\bf x}%
_{n}=( x_{0,n},x_{1,n}) $, where $x_{0,n}+x_{1,n}=1$
and $x_{k,n}\geq 0$ for $k=0,1$ and all $n$. These seven two-dimensional vectors are labeled by the
message number $n$ because their contents are updated every time the DLM
receives a message.
Note that the DLM stores information about the last message only.
The information carried by earlier messages is overwritten by
updating the internal registers.

Upon receiving the $(n+1)$th input event, the DLM performs the following steps:
It stores the first two elements of message ${\bf y}_{k,n+1}$ in
its internal register ${\bf Y}_{k,n+1}^{H}=(
C_{k,n+1}^{H},S_{k,n+1}^{H}) $, the middle two elements
of ${\bf y}_{k,n+1}$ in
${\bf Y}_{k,n+1}^{V}=( C_{k,n+1}^{V},S_{k,n+1}^{V})$,
and the last two elements of ${\bf y}_{k,n+1}$
in
${\bf Y}_{k,n+1}^{P}=( C_{k,n+1}^{P},S_{k,n+1}^{P})$.
Then, it updates its internal vector according to the rule~\cite{RAED05d}
\begin{equation}
x_{i,n+1}=\alpha x_{i,n}+( 1-\alpha ) \delta _{i,k},
\label{update}
\end{equation}%
where $0<\alpha <1$ is a parameter that controls the learning process~\cite{RAED05d}.
Note that by construction $x_{0,n+1}+x_{1,n+1}=1$, $x_{0,n+1}\geq 0$ and  $x_{1,n+1}\geq 0$.
From the solution of Eq.~(\ref{update}),
\begin{equation}
{\bf x}_n=\alpha^n {\bf x}_0 +(1-\alpha)\sum_{j=1}^{n-1}\alpha^{n-2-j}{\bf v}_j,
\end{equation}
the correspondence to the expression for the polarization in classical electrodynamics Eq.~(\ref{polarization})
can be seen. The vector ${\bf v}$ plays the role of the electric field vector ${\bf E}$
and the internal vector ${\bf x}$ plays the role of the polarization $P$.
Hence, one could say that the internal vector ${\bf x}$ is the response of the PBS to the incoming messages (photons)
represented by the vectors ${\bf v}$. Therefore the PBS "learns" so to speak from the information carried
by the photons. The characteristics of the learning process depend on the parameter $\alpha$
(corresponding to the response function).
Equation~(\ref{update}) is the simplest learning rule we could think of.
If experimental measurements for a single PBS would require another maybe more complicated rule to simulate the experimental outcome
then we could modify the learning rule but given the information we have right now Eq.~(\ref{update}) suffices.
}
\item{Transformation stage:
The second stage (T) of the DLM-based processor takes as input the
data stored in the six internal registers ${\bf Y}%
_{k,n+1}^{H}=( C_{k,n+1}^{H},S_{k,n+1}^{H}) $, ${\bf Y}%
_{k,n+1}^{V}=( C_{k,n+1}^{V},S_{k,n+1}^{V}) $, ${\bf Y}%
_{k,n+1}^{P}=( C_{k,n+1}^{P},S_{k,n+1}^{P}) $ and in the internal
vector ${\bf x}_{n+1}=( x_{0,n+1},x_{1,n+1}) $ and
combines the data into an eight-dimensional vector (see Fig.~\ref{PBS}).
Rewriting this vector as
\begin{equation}
\left(
\begin{array}{c}
\left( C_{0,n+1}^{H}+iS_{0,n+1}^{H}\right) C_{0,n+1}^{P}x_{0,n+1}^{1/2} \\
i\left( C_{1,n+1}^{V}+iS_{1,n+1}^{V}\right) S_{1,n+1}^{P}x_{1,n+1}^{1/2} \\
\left( C_{1,n+1}^{H}+iS_{1,n+1}^{H}\right) C_{1,n+1}^{P}x_{1,n+1}^{1/2} \\
i\left( C_{0,n+1}^{V}+iS_{0,n+1}^{V}\right) S_{0,n+1}^{P}x_{0,n+1}^{1/2}
\end{array}%
\right)
\equiv
\left(
\begin{array}{c}
a_{0}^{H} \\
ia_{1}^{V} \\
a_{1}^{H} \\
ia_{0}^{V}%
\end{array}%
\right) ,
\end{equation}
shows that the operation performed by the transformation stage T corresponds
to the matrix-vector multiplication in the quantum theoretical description
of a PBS, namely%
\begin{equation}
\left(
\begin{array}{c}
b_{0}^{H} \\
b_{0}^{V} \\
b_{1}^{H} \\
b_{1}^{V}%
\end{array}%
\right) =
\left(
\begin{array}{cccc}
1 & 0 & 0 & 0 \\
0 & 0 & 0 & i \\
0 & 0 & 1 & 0 \\
0 & i & 0 & 0%
\end{array}%
\right) \left(
\begin{array}{c}
a_{0}^{H} \\
a_{0}^{V} \\
a_{1}^{H} \\
a_{1}^{V}%
\end{array}%
\right),
\end{equation}%
where $(a_{0}^{H},a_{0}^{V},a_{1}^{H},a_{1}^{V})$ and
$(b_{0}^{H},b_{0}^{V},b_{1}^{H},b_{1}^{V})$ denote the input and output
amplitudes of the photons with polarization $H$\ and $V$ in the 0 and 1
channels of a PBS, respectively.
Note that in our simulation model there is no need to introduce the concept of a vacuum field,
a requirement in the quantum optical description of a PBS.
}

\item{Output stage: The final stage (O) sends the message
${\bf w}=\left(w_0,w_1,w_2,w_3,w_4,w_5\right)^T$,
where
\begin{eqnarray}
w_0 &=&C_{0,n+1}^{H}C_{0,n+1}^{P}\sqrt{x_{0,n+1}}/uw_4,\nonumber \\
w_1 &=&S_{0,n+1}^{H}C_{0,n+1}^{P}\sqrt{x_{0,n+1}}/uw_4,\nonumber \\
w_2 &=&-S_{1,n+1}^{V}S_{1,n+1}^{P}\sqrt{x_{1,n+1}}/uw_5,\nonumber  \\
w_3 &=&C_{1,n+1}^{V}S_{1,n+1}^{P}\sqrt{x_{1,n+1}}/uw_5,\nonumber \\
w_4 &=&\sqrt{w_0^{2}+w_1^{2}}/u,\nonumber \\
w_5 &=&\sqrt{w_2^{2}+w_3^{2}}/u,\nonumber\\
u &=&\sqrt{w_0^{2}+w_1^{2}+w_2^{2}+w_3^{2}},\nonumber \\
&&
\end{eqnarray}
through output channel 0 if $u^{2}>r$ where
$0<r<1$ is a uniform pseudo-random number.
Otherwise, if $u\le r$, the output stage sends
through output channel 1 the message
${\bf z}=\left(z_0,z_1,z_2,z_3,z_4,z_5\right)^T$,
where
\begin{eqnarray}
z_0 &=&C_{1,n+1}^{H}C_{1,n+1}^{P}\sqrt{x_{1,n+1}}/vz_4,\nonumber \\
z_1 &=&S_{1,n+1}^{H}C_{1,n+1}^{P}\sqrt{x_{1,n+1}}/vz_4,\nonumber \\
z_2 &=&-S_{0,n+1}^{V}S_{0,n+1}^{P}\sqrt{x_{0,n+1}}/vz_5,\nonumber \\
z_3 &=&C_{0,n+1}^{V}S_{0,n+1}^{P}\sqrt{x_{0,n+1}}/vz_5,\nonumber \\
z_4 &=&\sqrt{z_0^{2}+z_1^{2}}/v,\nonumber \\
z_5 &=&\sqrt{z_2^{2}+z_3^{2}}/v,\nonumber \\
v &=&\sqrt{z_0^{2}+z_1^{2}+z_2^{2}+z_3^{2}}.\nonumber \\
&&
\end{eqnarray}
}
\end{itemize}
Any other algorithm
that selects the output channel in a systematic manner might be employed as well.
This will change the order in which messages are being processed but the content
of the messages will be left intact and the resulting averages do not change significantly.

\subsection{Remaining optical components}
The Wollaston prism is a PBS with one input channel and two output channels
and is simulated as the PBS described earlier.

In contrast to the PBS, the HWP and the EOM are passive devices.
As can be seen from the wave mechanical description, a HWP
does not only change the polarization of the photon but also its phase~\cite{BORN64}.

When a voltage is applied to the EOM, $R\neq 0$ (see Eq. (2) in~\cite{JACQ08}) and the EOM acts as a wave plate that rotates the
polarization of the incoming photons by an angle depending on $R$. In the simulation a pseudo-random number is used
to decide to apply a voltage to the EOM or not.
Also here we use a pseudo-random number to mimic the experimental procedure to control the EOM~\cite{JACQ08,JACQ07}.
Any other (systematic) sequence to control the EOM can be used as well.

\subsection{Detection and data analysis procedure}Detector $D_0$ ($D_1$) registers
the output events at channel 0 (1).
During a run of $N$ events, the algorithm generates the data set
\begin{equation}
\Gamma (R)=\left\{x_{n},y_{n},A_{n}|n=1,...,N;\Phi =\Phi_{1}-\Phi _{0}\right\} ,
\end{equation}
where $x_{n}=0,1$ indicates which detector fired ($D_{0}$ or $D_{1}$),
$y_{n}=0,1$ indicates through which arm of the MZI the messenger (photon)
came that generated the detection event (note that $y_{n}$ is only
measured in the simulation, not in the experiment),
and $A_{n}=0,1$ is a pseudo-random number that is chosen
after the $n$th messenger has passed the first PBS,
determining whether or not a voltage is applied to the
EOM (hence whether the MZI configuration is open or closed).
Note that in one run of $N$ events a choice is made between no voltage or a particular
voltage corresponding to a certain reflectivity $R$ of the output BS
(see Eq. (2) in~\cite{JACQ08}).
The angle $\Phi $ denotes the phase shift between the two interferometer arms.
This phase shift is varied by applying a plane rotation on the phase of the particles entering
channel 0 of the second PBS. This corresponds to tilting the second PBS in
the laboratory experiment~\cite{JACQ08}.
For each $\Phi $ and
MZI configuration the number of 0 (1) output
events $N_{0}$ ($N_{1}$) is calculated.

\setlength{\unitlength}{1cm}
\begin{figure}[t]
\begin{center}
\includegraphics[width=7.5cm]{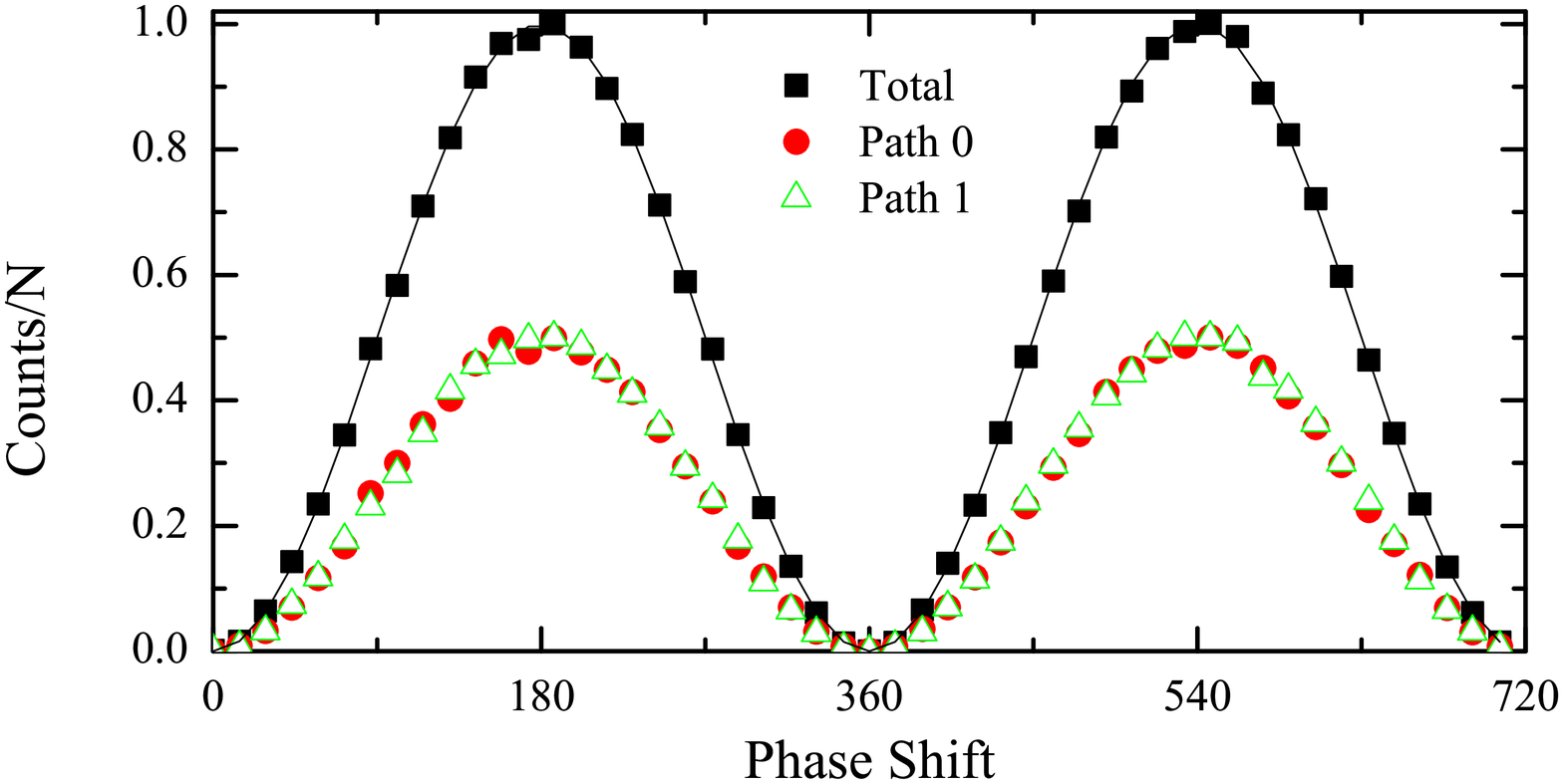}
\caption{(Color online) Event-by-event simulation results of the interference visibility $V$
for $R=0.5$.
Markers give the results for the normalized intensity $N_0/N$
as a function of the phase shift $\Phi$, $N_0$
denoting the number of events registered at detector D$_0$.
Circles (triangles) represent the detection events generated by photons that followed path 0 (1)
and squares represent the total number of detection events.
For each value of $\Phi$,
the number of input events $N=10000$.
The total number of detection events per data point (squares) is approximately the same as in experiment.
The solid line represents the results of quantum theory.
}
\label{malus}
\end{center}
\end{figure}

\setlength{\unitlength}{1cm}
\begin{figure*}[t]
\begin{center}
\includegraphics[width=15cm]{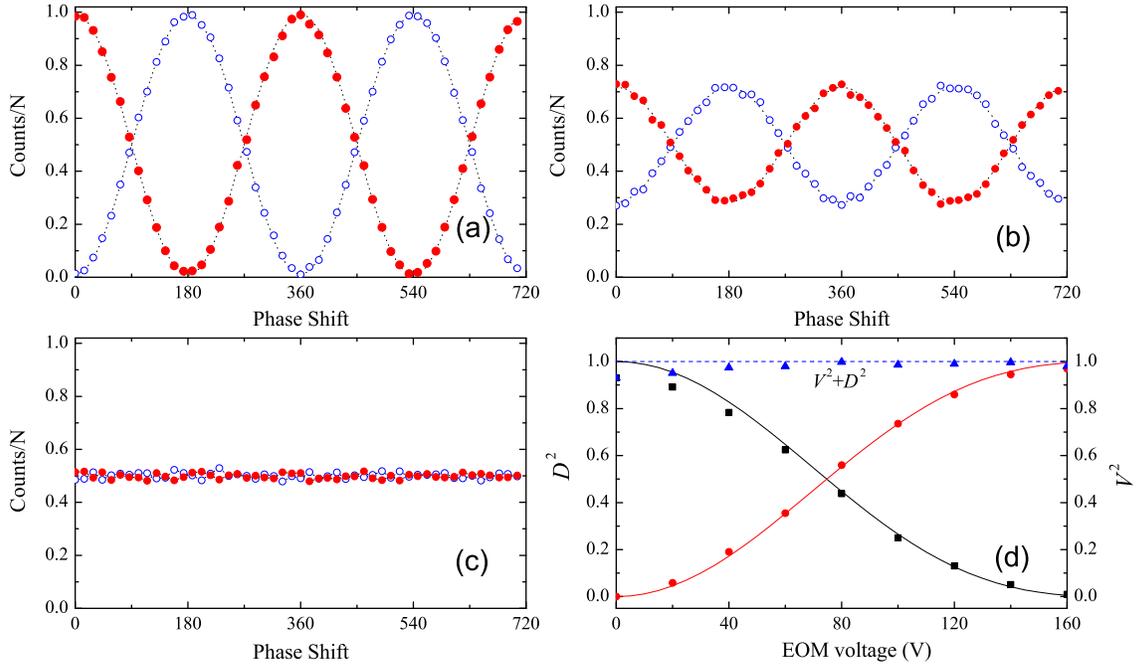}
\end{center}
\caption{(Color online) Event-by-event simulation results of the interference visibility $V$
for different values of $R$ ((a)-(c)) and of $V^2$, $D^2$ and $V^2+D^2$ as a function of the EOM voltage (d).
(a)-(c) Circles give the results for the normalized
intensities $N_{0}/N$ and $N_{1}/N$ as a function of the phase shift $\Phi$, $N_{0}$ ($N_{1}$) denoting
the number of events registered at detector $D_{0}$ ($D_{1}$), for
(a) $R=0.43$ ($V\approx 0.98$), (b) $R=0.05$ ($V\approx 0.45$) and (c) $R=0$ ($V=0$).
For each value of $\Phi$, the number of input events $N=10000$. The number
of detection events per data point is approximately the same as
in experiment. Dashed lines represent the results of quantum theory (Malus law).
(d) Circles give the simulation results. Lines represent the theoretical expectations obtained from Eqs.~(2), (3) and (7)
in~\cite{JACQ08} with $\beta=24^o$ and $V_{\pi}=217 V$.
}
\label{fig}
\end{figure*}

\section{Simulation results}
The algorithm described above directly translates
into a simple computer program that simulates the messenger routing
in a network that contains all the optical components of the
laboratory experiment~\cite{JACQ08}.
Before the simulation starts we set ${\bf x}%
_{0}=( x_{0,0},x_{1,0}) =( r,1-r) $, where $r$ is a
uniform pseudo-random number. In a similar way we use pseudo-random numbers
to initialize ${\bf Y}_{0,0}^{H}$, ${\bf Y}_{0,0}^{V}$, $%
{\bf Y}_{0,0}^{P}$, ${\bf Y}_{1,0}^{H}$, $%
{\bf Y}_{1,0}^{V}$ and ${\bf Y}_{1,0}^{P}$.
In this simulation, we send messengers to one input channel of the input PBS only
(see Fig.~\ref{wheeler}).
The HWP in BS$_{output}$ changes the phases and also interchanges the roles of channels 0 and 1.
Disregarding a few exceptional events, the PBS in BS$_{output}$ generates messages in one of the channels only.
For a fixed set of input parameters, each simulation takes a few seconds on a
present-day PC.
In all simulations, $\alpha=0.99$~\cite{RAED05d}.

We first demonstrate that our model yields full WPI of the photons.
Figure~\ref{malus} shows the number of detection events at $D_0$ as a function of $\Phi$ for $R=0.5$.
The events generated by photons following path 0 and path 1 of the MZI are counted separately.
It is clear that the number of photons that followed path 0 and path 1 is equal and that the total intensity in output channel 0
obeys Malus law. Hence, although the photons have full WPI for all $\Phi$ they can build an interference pattern by arriving one by one at a detector.

Next, we calculate for $R=0,0.05,0.43$~\cite{JACQ08} and for each phase shift $\Phi$ and configuration (open or closed) of the MZI the number of
events registered by the two detectors behind the output BS, just like in the experiment.
Figure~\ref{fig}(a)-(c) depicts the interference visibility $V$.
The simulation data quantitatively agree with the averages calculated from quantum theory and
qualitatively agree with experiment (see Fig.3 in \cite{JACQ08}).
Calculation of $D$ as described in \cite{JACQ08} gives the results for $D^2$ and $V^2$ shown in Fig.~\ref{fig}(d).
Comparison with Fig.4 in \cite{JACQ08} shows excellent qualitative agreement.

\section{Conclusion}

In this paper, we have presented a simulation model that is solely based on experimental facts,
that satisfies Einstein's criterion of local causality, that does not rely
on any concept of quantum theory or of probability theory,
and that provides a description of the experimental observations in~\cite{JACQ08} on the level of
individual events.
In a pictorial description of our simulation model, we may speak about ``photons'' generating the detection events.
In the simulation we can always track the photons, even in the closed configuration of the MZI. The photons always have full WPI,
never directly communicate with each other, arrive one by one at a detector but nevertheless build up an interference pattern at the detector in
the case of the closed configuration of the MZI.
Hence, although for $0<R\le 0.5$ we find that $0\le D< 1$ and $D^2+V^2\leq 1$ with values for $D$ and $V$ in qualitative agreement
with the experimental results, we always have access to full WPI, even in the case $D=0$, $V=1$.
Our model thus provides a counter example for the fact that full WPI would correspond to $D=1$.
A further consequence is that the relation $V^2+D^2\le 1$ cannot be regarded as quantifying the notion of complementarity: Our model
allows a particle-only description for both the open and closed configuration of the MZI.

\bibliographystyle{elsarticle-num}

\end{document}